\documentstyle[psfig,conf_iap,10pt]{article}

\newcommand{\lya}{Ly$\alpha$}
\newcommand{\kms}{~{\rm \ km \ s}^{-1}}

\begin{document}
\heading{%
%
 \lya\ absorbers arising in galaxy clusters
%
} 
\par\medskip\noindent
\author{%
A. Ortiz-Gil$^{1,2}$, K.M. Lanzetta$^{2}$, J.K. Webb$^{1}$, 
X. Barcons$^{3}$
}
\address{%
Dept.\ of Astrophysics \& Optics, UNSW, Sydney 2052, NSW, Australia
}
\address{%
Dept.\ of Physics \& Astronomy, SUNY, Stony Brook, NY 11794-3800, USA 
}
\address{%
IFCA, CSIC-UC, Avda. de los Castros s/n, E-39005, Santander, Spain
}

\begin{abstract}
We present here new GHRS observations of \lya\ absorption lines associated 
with groups or clusters toward the QSOs 1545+2101 and 0850+4400.
In the first case we have identified at least
eight distinct Ly$\alpha$ absorption features, with a mean redshift of
$<z> = 0.2648 \pm 0.0002$ and a velocity dispersion of
$163 \pm 57 \kms$. We have also identified a group or cluster of
galaxies in the vicinity of this QSO with a mean redshift of $<z> =
0.2643 \pm 0.0004$ and velocity dispersion $223 \pm 91 \kms $.
The spectrum of QSO 0850+4400, of poorer quality, reveals two \lya\ absorption
features at $z=0.0909500\pm 0.0000070$ (which is just resolved) and
$z=0.0948215\pm 0.0000090$, separated by $\sim 1060\, \kms$. A group 
or cluster of galaxies is also present in the vicinity of the QSO line-of-sight
with a mean redshift $<z>=0.0901\pm 0.0007$ and velocity dispersion of
$530 \pm 200 \kms $. The results of this work establish that \lya\
absorption can occur in denser than average galaxy environments, and that 
it arises in discrete objects spanning a velocity range similar to that of the
cluster galaxies. Although a one-to-one relationship between absorbers 
and galaxies is difficult to establish in such a dense environment,
the results obtained here are indeed consistent with the \lya\ absorption lines
being associated with individual galaxies also in groups and clusters. 
Moreover, the data shows clearly that line clustering takes place in the \lya\
forest. 

\end{abstract}


\section{Absorption systems \& groups of galaxies}
Details about the GHRS spectra of 1545$+$2101 and 0850$+$4400 
(see Fig. 1) will be presented elsewhere \cite{ortiz}. A standard Voigt 
profile fitting \cite{lanzetta1} to individual lines (or groups of them) 
shows the following features in each field:
(a) 1545+2101: two lines are found at $z=0.2504707 \pm 0.0000030$ and 
$z=0.2522505 \pm 0.0000016$, with a velocity separation of $\sim 427\, \kms$.
Also present is a group of eight lines, whose redshift centroid is 
$<z>= 0.2648 \pm 0.0002$ with a velocity dispersion of $163 \pm 57 \kms$.
(b) 0850+4400: two lines are detected at $z=0.0909500 \pm 0.0000070$ and 
$z=0.0948215 \pm 0.0000090$, with a velocity separation of $1062 \pm 3\, \kms$.

A number of galaxies in these two fields have been observed spectroscopically
\cite{lanzetta2} to determine their possible connection to the absorption 
features detected on the QSOs spectra:
(a) 1545+2101: a galaxy is found at $z=0.2510$, its impact parameter value 
being $\rho=306.4 \ h^{-1}$ kpc. Moreover, there is 
a group or cluster of six galaxies with redshift centroid
$<z>=0.2643 \pm 0.0004$ and velocity dispersion of $223\pm 91 \kms $, whose 
impact parameters range from $47$.6 to $456.4  h^{-1}$ kpc.
(b) 0850+4400: there is a group or cluster of seven galaxies with redshift 
centroid $<z>=0.0901 \pm 0.0007$ and velocity dispersion of $530 \pm 200
\kms $. Impact parameter range is $38.5$ -- $117.8 \ h^{-1}$ kpc 
($h~=~H_0/(100 \kms \mbox{Mpc}^{-1})$, $q_0=0.5$).

\begin{figure}
\centerline{\hbox{
\psfig{figure=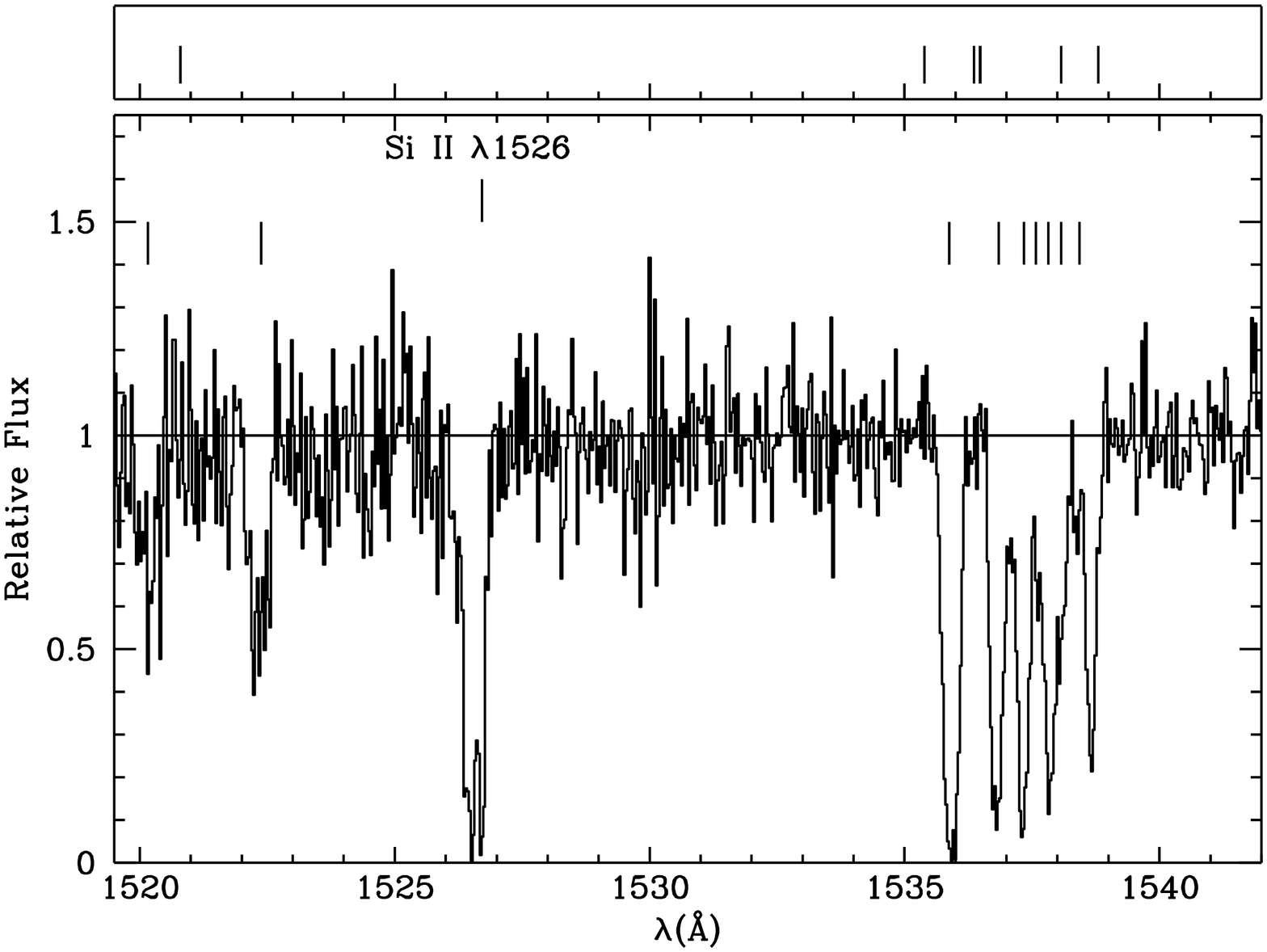,height=3.6cm,angle=-90}
\psfig{figure=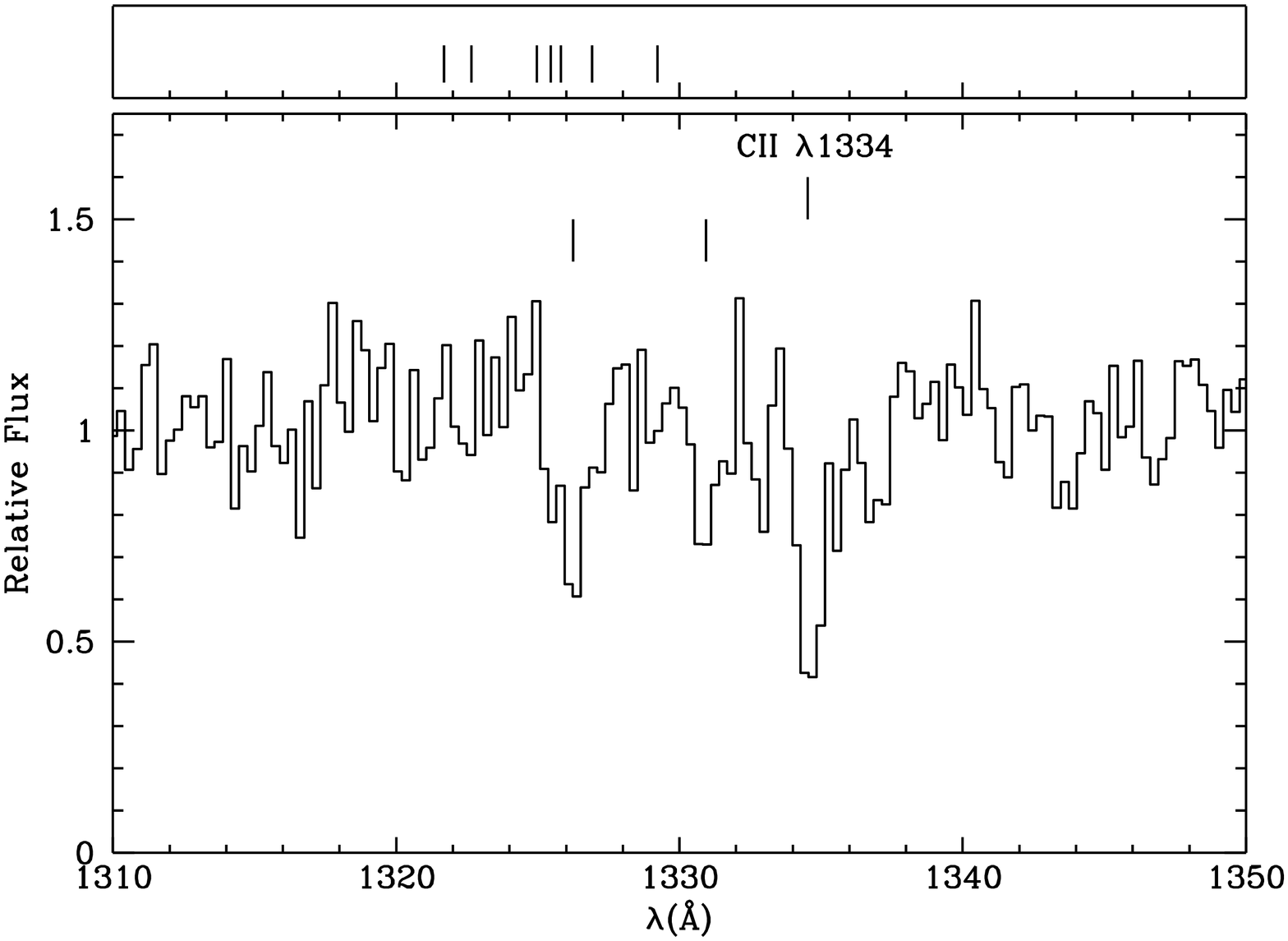,height=3.6cm,angle=-90}
}}
\caption[]{
Spectrum of 1545+2101 (left) and enlarged extract from spectrum of
0850+4400 (right). Tick marks in the upper 
pannel indicate the predicted wavelengths of \lya\ at the redshifts of the 
galaxies. Tick marks in the lower pannel show the positions of the detected 
absorption lines}
\end{figure}

The fact that we observe a number of resolved lines shows that the 
absorptions arise in discrete clouds of cold gas instead of arising in a 
large single diffuse gas component. 
The comparison between the cross-correlation function (CF) between the galaxies
and absorbers in the data with the CF corresponding to a random case shows 
that there is a non-random connection between the absorbers and the galaxies 
in our data. If we assume that galaxies within the real group or cluster are 
distributed in velocity space according to a gaussian distribution, a 
one-to-one match between the absorbers and the galaxies cannot be established
with the present data.

\section{Conclusions}

From these results we can conclude the following: (1) The detection 
of groups of \lya\ absorption lines implies that some of the \lya\ absorbers 
do actually cluster.
(2) The velocity spanned by the \lya\ absorption lines arising in
groups or clusters is consistent with the velocity dispersion of the
corresponding group or cluster of galaxies. This implies that the
\lya\ absorbers arising in those clusters occupy the same region in
space than the galaxies themselves.
(3) There is no strong preference for \lya\ absorbers to avoid overdense
environments.
(4) \lya\ absorption caused by galaxy groups or clusters arises in
discrete clouds of cold gas, rather than in some hypothetical smoothly
distributed cold phase in the intracluster medium.

\vfill
\end{document}